# Chapter 8: Cool Pavements

Martin Hendel[1,2]

[1] Univ Paris Diderot, Sorbonne Paris Cité, LIED, UMR 8236, CNRS, F-75013, Paris, France
[2] Université Paris-Est, ESIEE Paris, département SEN, F-93162 Noisy-le-Grand, France

## Abstract

Cool pavements designate alternative pavements designed to reduce their contribution to urban heating. Urban heating generally refers to the sensible heat exchanged with the atmosphere by urban materials but can also include the radiative load they impose on pedestrians. In either case, pavement surface temperature is a very important parameter, which cool pavements seek to decrease compared with standard pavement designs.

The energy balance of a pavement surface or very thin pavement slab helps identify the outbound flows which cool pavements attempt to promote and fundamental physical principles which govern them. On this basis, cool pavements can be classified as reflective pavements, green and evaporative pavements, high inertia or phase changing pavements as well as conductive or heat-harvesting pavements.

This chapter presents the urban heat island and the urban heating phenomena and provides an overview of cool pavement technologies, detailing areas which require further scientific investigation.

## Nomenclature

| | |
|---|---|
| $\alpha$ | albedo |
| $\epsilon$ | emissivity |
| $E$ | evaporation rate |
| $H$ | sensible (convective) heat flux density [W.m$^{-2}$] |
| $l$ | latent heat of vaporization of water [kJ.kg$^{-1}$] |
| $\lambda$ | thermal conductivity [W.m$^{-1}$.K$^{-1}$] |
| $L$ | LW radiation [W.m$^{-2}$] |
| LW | longwave (3-100 μm) |
| NIR | near infrared (0,8-3 μm) |
| PCM | phase change material |
| $p_0$ | atmospheric pressure [Pa] |
| $p_S$ | water vapor saturation pressure at $T_S$ [Pa] |
| $p_v$ | ambient water vapor pressure [Pa] |
| PV | photovoltaic |
| $R_n$ | net radiation [W.m$^{-2}$] |
| $\sigma$ | Boltzmann constant [W.m$^{-2}$.K$^{-4}$] |
| $S$ | SW radiation [W.m$^{-2}$] |
| SW | shortwave (0,3-3 μm) |
| $T_z$ | pavement temperature at depth z [°C] |
| $T_{air}$ | air temperature [°C] |
| $T_s$ | surface temperature |
| TEG | thermoelectric generator |
| UHI | urban heat island |
| UHII | urban heat island intensity |
| $z$ | depth [m] |

## 8.1. Introduction

In response to the public health-threat posed by global warming and heat-waves, cities around the world are looking for adaptation solutions to limit the impacts of these global challenges. Among other solutions, so-called "cool pavements" have been proposed and studied for decades with the aim of countering urban heat islands.

The term "cool pavement" is loosely defined, generally referring to pavements engineered to limit their surface temperature and thus their contribution to the urban heat island effect compared to standard materials. Cool pavements belong to the broader family of cool urban materials which also include building rooftops and façades.

Initially, cool materials mostly consisted of white or light-colored coatings and materials. Work on this topic began decades ago with high-albedo roofing materials, with a number of papers focusing to determine their potential to reduce UHII (Akbari and Taha, 1992; Taha, 1997; Akbari, Pomerantz and Taha, 2001). More recently, cool-colored materials have been developed which are highly reflective in the near infrared (NIR) band, spanning from 0.8 μm to 3 μm in wavelength. This allows for a wide variety of colors while still providing cooling performance (Santamouris, Synnefa and Karlessi, 2011). In addition to reducing urban temperatures, SW reflective materials create a small one-off negative radiative forcing that counters the effects of increasing greenhouse gas concentrations (Akbari and Matthews, 2012). Similar solutions exist in the case of pavements.

In the case where materials have low emissivity, such as metallic roofing materials, high emissivity coatings can be used to increase LW radiant cooling (Météo France and CSTB, 2012). Due to their relatively high emissivity, this kind of approach is not frequently found in pavements.

Another category of cool material are so-called green materials. These are typically applied to building roofs and façades or street surfaces and rely on the use of green walls and roofs or lawn surfaces (Akbari, Pomerantz and Taha, 2001; Wong *et al.*, 2009; Ng *et al.*, 2012). Indeed, while the benefits of lawns, parks and trees have been studied and promoted for several decades (Jauregui, 1990; Taha, 1997), the more recent green facades and roofs make it possible to green both horizontal and vertical building surfaces (Jaffal, Ouldboukhitine and Belarbi, 2012; Musy *et al.*, 2014; Djedjig, Bozonnet and Belarbi, 2016). In addition to cooling cities, green materials have other benefits such as promoting urban biodiversity and reducing rainwater runoff. Green pavement solutions can be found for low-traffic applications.

For high traffic surfaces which cannot be vegetated, permeable paving materials have been developed (Kubo, Kido and Ito, 2006; Haselbach *et al.*, 2011; Li *et al.*, 2013). These mainly consist of concrete or asphalt concrete with high void content which allows water to drain into the sublayers of the pavement (Li *et al.*, 2013).

Another way to decrease the surface temperature of urban materials is to use the latent heat storage properties of a phase change material (PCM) to increase the material's thermal inertia (Santamouris, Synnefa and Karlessi, 2011). Such applications can also be found in pavements.

Cool pavements are of growing interest to cities across the world which have broadly adopted the goal of countering the combined effects of heat waves and urban heat islands (UHI). This awareness is accelerated by climate change forecasts, which predict an increase in these phenomena and their impacts, with certain cities even facing the risk of becoming uninhabitable all or part of the year (Pal and Eltahir, 2015; Mora *et al.*, 2017; Kang and Eltahir, 2018). These challenges have spurred the interest of urban decision-makers and planners for cooling techniques including cool pavements, consequently stimulating research in the field.

The present chapter seeks to present an overview of the most frequent cool pavement designs and the physical processes their performance is based on. To this aim, section 8.2 will briefly present the UHI effect with a focus on the urban energy budget. Section 8.3 will present an overview of different cool pavement technologies and describe their future research needs.

## 8.2. The urban heat island effect and the urban energy balance

### Urban heat island effect

Urban areas, through a combination of radiative trapping, increased heat storage, wind obstruction, reduced presence of vegetation, low surface permeability, as well as high concentrations of human activity, create a localized warming effect known as the UHI effect (Oke, 1982; Grimmond, 2007). The UHI effect is observed as increased urban air and surface temperatures compared with surrounding rural areas, in the order of 1-3°C on average (Akbari, Pomerantz and Taha, 2001). The air temperature difference, measured as that of the urban area minus that of the rural area, is referred to as urban heat island intensity (UHII).

UHII varies with time and local weather conditions. Typically, high wind speeds and cloud cover tend to decrease it, while clear skies and low winds amplify it (Cantat, 2004). On nights with clear skies and calm winds, UHII can reach as much as 12°C (Oke, 1973). These conditions are typically reached during heat-waves.

While UHI may reduce building energy consumption during winter, the opposite is true in the summer during which cooling demand is increased (Hassid *et al.*, 2000; Santamouris, 2014). For Athens, Greece for example, building cooling energy has been found to double when UHII reaches 10°C, while peak cooling electricity demand is tripled (Santamouris *et al.*, 2001).

In addition to negative effects on energy use, UHIs also tend to exacerbate ozone and smog pollution (Rosenfeld et al., 1998) as well as the intensity of heatwaves (Li and Bou-Zeid, 2013). As a result, their health impacts are increased, making heat-waves of particular concern for dense urban areas with intense UHIs, as seen in Europe during the 2003 heat-wave, particularly devastating in Paris, France (Robine *et al.*, 2008).

Luke Howard, frequently referred to as the father of meteorology and the namer of clouds, was the first scientist to formally observe this phenomenon, keeping a strict record of air temperature measurements between his home in the outskirts of 19[th] Century London and the Royal Society building in the city center (Howard, 1833). Since then, knowledge and understanding of the UHI phenomenon has gradually shifted from empirical descriptions and predictions (Oke, 1973), towards a finer understanding of the physical processes involved (Oke, 1982). Currently, several urban canopy models have been developed and are able to represent the effects of low-level vegetation, green roofs, trees and even urban hydrology (Masson, 2000; Kusaka, Kondo and Kikegawa, 2001; Dandou *et al.*, 2005; Grimmond *et al.*, 2011). These tools have proved invaluable to evaluate the impact of different urban cooling scenarios at the city-scale (Daniel, Lemonsu and Viguié, 2016).

### The urban energy budget

Two principal scales can be considered when analyzing the urban energy balance: that of an urban volume or an urban surface or facet.

### Urban volume energy balance

Figure 1 illustrates the energy balance of an urban volume whose upper limit is above the urban canopy layer and whose depth is such that the conductive heat flux is negligible over the time scale considered.

The term $H$ represents the convective exchange between the ground and the atmosphere; $IE$ is the latent heat flow due to the (possible) evaporation of water present at the surface. The latter is the product of the latent heat of water vaporization $l$ and the evaporation rate $E$. Net downwards radiation is noted $R_n$. $R_n$ is composed of four components (see Figure 2): $S$ and $L$ for incident short and long wavelengths respectively (SW and LW) and $S_{up}$ and $L_{up}$ for SW and LW radiosity, respectively, summarized in the following equation:

$$R_n = S + L - L_{up} - S_{up} R_n = S + L - L_{up} - S_{ref} \qquad (1)$$

In addition, $Q_F$ refers is the atmospheric anthropogenic source term, i.e. waste heat due to human activities; $\Delta Q$ is the heat storage term within the urban materials present in the volume. The term $Q_A$ represents non-atmospheric heat advection outside of the volume. This exchange can be horizontal, for example heat transport by a river, but can also be vertical, e.g. in the case of geothermal energy. Atmospheric heat advection, when the area around the urban volume is homogeneous, is included in the term $H$.

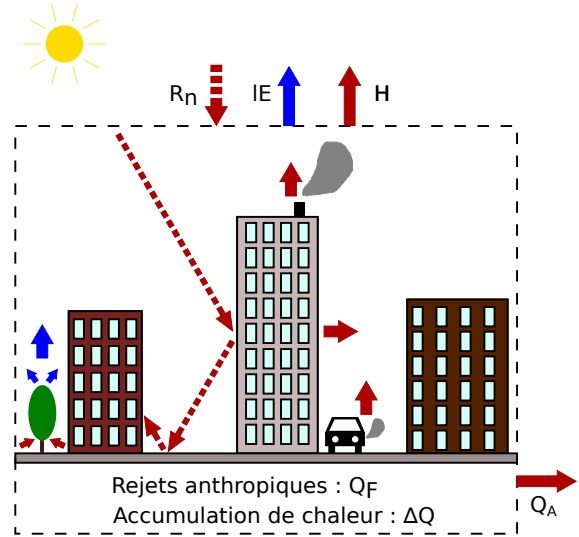

*Figure 1: Urban volume energy balance*

The energy conservation equation for an urban volume is:

$$R_n + Q_F = H + lE + \Delta Q + Q_A \qquad (2)$$

**Pavement surface energy balance**

The urban energy balance of a pavement surface is shown in Figure 2.

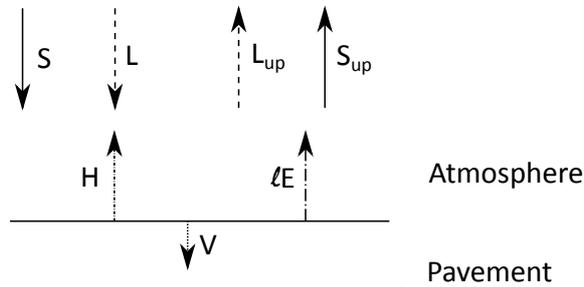

*Figure 2: Pavement surface energy balance*

The term $V$ represents surface conduction into the pavement. This component is apparent at this scale and was internalized in the term $\Delta Q$ in equation (2) when considering an urban volume.

The energy balance of an urban facet is given by:

$$R_n = H + lE + V \qquad (3)$$

Pavement temperature, which influences each of the terms in the previous equation, automatically varies to compensate any imbalance that may arise.

To illustrate the relative importance of these terms, Table 1 and Table 2 provide orders of magnitude of their intensity for a mid-latitude city under clear skies in summer, with Table 2 focusing on the terms of the radiative balance.

*Table 1: Clear sky summertime heatflows for a mid-latitude city in W/m² (Oke, 1988).*

|       | $R_n$ | $Q_F$ | $lE$ | $H$ | $\Delta Q$ |
|-------|-----|-----|-----|-----|-----|
| Day   | 516 | 30  | 158 | 240 | 148 |
| Night | -80 | 20  | 13  | 7   | -80 |

*Table 2: Clear sky summertime radiative balance for a mid-latitude city in W/m² (Oke, 1988).*

|       | $S$ | $L$ | $S_{up}$ | $L_{up}$ |
|-------|-----|-----|-----|-----|
| Day   | 760 | 365 | 106 | 503 |
| Night | 0   | 335 | 0   | 415 |

Clearly, the most important incoming flux is the net radiation $R_n$, more specifically solar irradiance $S$. Compared to a natural area, the main differences in the energy balance in the city are the positive values of day and night $H$ and the importance of the storage term $\Delta Q$ at night (Oke, 1988). In addition, latent flows are lower in cities than in the countryside, while convective flows are higher.

When considering the heat balance of a pavement surface, it is more appropriate to expand the radiative term such as to explicitly present inbound and outbound flows in equation (3):

$$S + L = S_{up} + L_{up} + H + lE + V \qquad (4)$$

In this form, the terms on the left are the inbound flows, while the terms on the right represent the outbound flows. From the perspective of pavement design, the former are fixed by site selection and the presence or absence of radiative masks, while the terms on the right are dependent on the pavement's thermos-physical properties, as detailed below.

As mentioned in the Introduction, the goal of cool pavements is to limit their contribution to urban warming compared to standard materials. In other words, they seek to modify the surface energy balance so as to reduce the term $H$ in equation (3). This term is governed by Jürges' (1924) formula:

$$H = h(T_S - T_{air}) \qquad (5)$$

The parameter $h$ is the convective heat exchange coefficient, which is dependent on the air circulation conditions. $T_{air}$ is the ambient air temperature, while $T_S$ is the surface temperature. Given that $h$ and $T_{air}$ are fixed parameters from the perspective of the choice of pavement material, reducing the term $H$ is equivalent to reducing its surface temperature.

Detailing the other terms in equation (4) provides the following set of equations:

$$S_{up} = \alpha S \qquad (6)$$
$$L_{up} = (1 - \epsilon)L + \epsilon \sigma T_S^4 \qquad (7)$$

$$V = -\lambda \frac{\partial T_z}{\partial z}\bigg|_{z=0} \quad (8)$$

$$lE = 0.622 \frac{lh}{c_p p_0} T_S \left(\frac{p_s}{T_S} - \frac{p_v}{T_{air}}\right) \quad (9)$$

Equations (5) and (6) relate the physical laws which define SW and LW radiosity, respectively. As can be seen, SW radiosity is solely due to SW reflection, directly governed by the pavement's surface albedo α. Furthermore, LW radiosity depends both on LW reflectivity and LW emission as governed by the Stefan-Boltzmann law. The latter parameter is dependent on surface temperature and emissivity ϵ.

Equation (7) expresses the Fourier conduction law under 1D conditions in the vertical direction, mainly impacted by the vertical temperature profile $T_z$ and the material's thermal conductivity λ.

Finally, equation (8) governs the latent heat flux for a water film with $c_p$ the specific heat of air and $p_0$, $p_s$ and $p_v$ representing total air pressure, saturation vapor pressure at the water film temperature $T_S$ and partial air vapor pressure at $T_{air}$, respectively. A simple description of its physical meaning states that the evaporation rate increases with the water film temperature, principal driver of the vapor pressure deficit. For better clarity, the latent flow is left in its initial form in equation (9), obtained by inserting equations (5) through (8) into equation (4):

$$S + L = \alpha S + (1 - \epsilon)L + \epsilon \sigma T_S^4 + h(T_{air} - T_S) + lE - \lambda \frac{\partial T_z}{\partial z}\bigg|_{z=0} \quad (9)$$

As mentioned in the previous section, the terms on the right depend on the pavement's properties, more specifically on its albedo α, emissivity ϵ and thermal conductivity, in addition to the presence of water in or on the pavement to permit evaporation. It is by acting on these properties that cool pavements can offer improved thermal performance in cities. In addition, surface temperature is a fundamental parameter.

Modelling this last parameter more fully requires a slight modification of the previous heat budget, obtained by considering a thin layer of pavement of depth *dz* as illustrated in Figure 3.

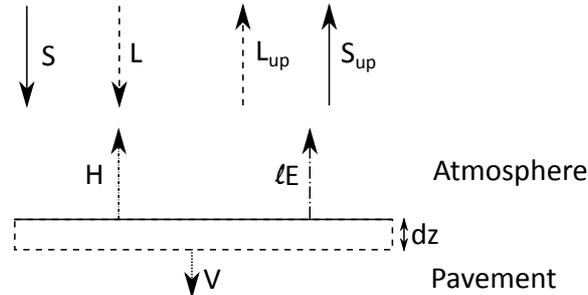

*Figure 3: Energy balance of a pavement slab of thickness dz.*

The following equation is then obtained, with *C* the specific heat and ϱ the density of the pavement material:

$$S + L = \rho C \frac{\partial T_S}{\partial t} dz + \alpha S + (1 - \epsilon)L + \epsilon \sigma T_S^4 + h(T_{air} - T_S) + lE - \lambda \frac{\partial T_z}{\partial z}\bigg|_{z=dz} \quad (10)$$

This differential equation illustrates the role of pavement heat capacity on regulating the thermal power of the heat flows in equation (4).

Having explained the fundamental physical laws that govern pavement thermal behavior, we now proceed to analyze specific kinds of cool pavements: reflective, green and evaporative, high inertia, and high conduction and heat-harvesting pavements.

## 8.3. Overview of cool pavement technologies

We begin by focusing on reflective pavements.

**Reflective pavements**

In order to reduce their surface temperature, reflective pavements focus on limiting their absorption of solar energy, principally by an increased albedo compared to standard pavements.

Albedo is defined as the proportion of radiation reflected by a horizontal surface. This parameter is dependent on the spectral composition of the incident radiation. On Earth, albedo is determined according to the spectral composition of solar radiation. Usually, the spectral composition of global irradiance for air mass 1.5 (AM1.5) at sea level is considered, available from ASTM G173 and illustrated in Figure 4. The dashed vertical lines delimit the visible light spectral range (380-740 nm).

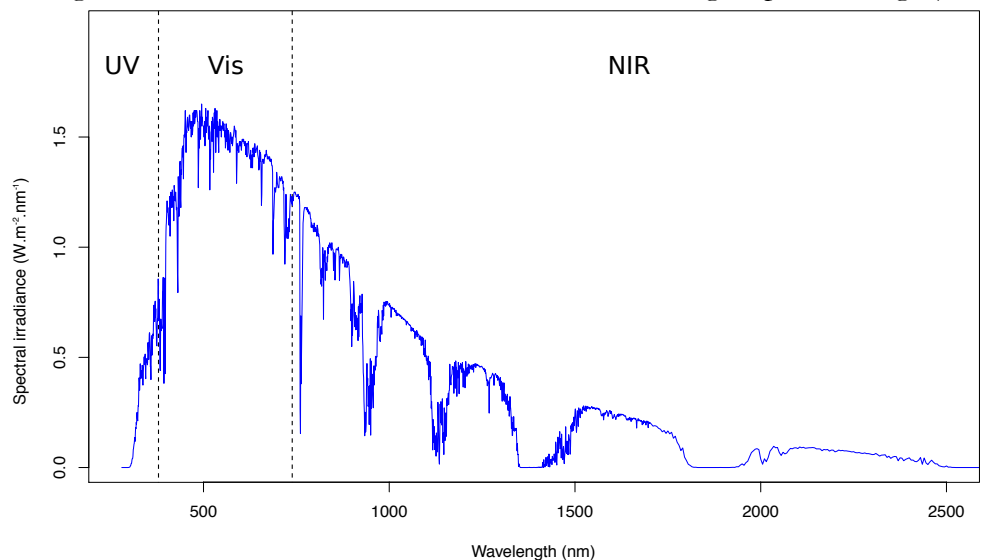

*Figure 4: Standard solar spectral irradiance for air mass 1.5 (AM 1.5)*

Two principal ASTM standards are used to measure albedo: ASTM E903 for lab measurements using a UV-VIS-NIR Spectrophotometer and integrating sphere (ASTM E903-12, 2015), or ASTM E1918 for measurements in the field using a pyranometer or an albedometer (ASTM E1918-16, 2016).

Solar energy at sea level is partitioned as follows: 3% in the ultraviolet (UV) range, 49% in the visible (VIS) range, and 46% in the near-infrared (NIR) range. Consequently, engineering efforts focus on increasing the reflectivity of pavement materials in the visible and near-infrared range.

There are two types of reflection: specular and diffuse. Specular reflection is akin to that of a mirror, i.e. incident light is reflected in a specific direction only. Diffuse reflection is akin to that of a matte white painted wall, i.e. reflected light is scattered in all directions. In reality, most materials display a combination of both behaviors which may vary according to the angle of incidence. Typically, asphalt concretes display mainly diffuse reflection at close-to-normal angles of incidence, and specular reflection for grazing angles of incidence.

Standard paving materials span a range of possible albedo values. Asphalt pavements have the lowest values, typically 0.05 for new asphalt concrete, which increases with aging and soiling, reaching up to 0.15. Concrete pavements are brighter and can reach an albedo of up to 0.35 for standard Portland cement concretes. On the contrary, their albedo decreases with time due to soiling, typically reaching 0.2 within a few weeks or months following installation. For pavers, typically used for pedestrian applications, their albedo varies with their composition. Reflective concrete pavements have been shown to reach albedos of up to 0.77, while cool coatings applied to asphalt concrete can have an initial albedo above 0.5.

For asphalt surfaces, aggregate albedo is an important factor in pavement albedo, while for concrete surfaces, this depends greatly on cement albedo. This is caused by the structure of both pavements and how they age with wear. While road traffic tends to wear away the bituminous binder, which has low albedo, from asphalt concrete, thus exposing its aggregates, cement is not worn off from concrete. Its aggregates are therefore less exposed, even after road wear. On the other hand, the cement becomes soiled and stained from road traffic.

To increase the albedo of a pavement, several solutions have been developed. These range from the use of light-colored binders or aggregates to deploying a reflective coating or thin surface layer for asphalt pavement structures (Santamouris, 2013). The added layer can be composed of concrete, in which case the process is known as "whitetopping". Light-colored aggregate chip seals can also be used (Rosado *et al.*, 2014). Alternative designs can use a resin-based binder and alternative aggregates (Anting *et al.*, 2018). For concrete, white cement can be used although its production is significantly more energy-intensive than regular gray cement. Alternatively, concrete doped with $TiO_2$ or slag cement can also be used. The surface layer of both asphalt and concrete pavements can also be painted over with a reflective coating.

To limit the glaring effect of reflective pavements for pedestrians or drivers, cool-colored pavements have been proposed, such as a NIR-reflective paint coating for pavements studied experimentally by Kinouchi *et al.* (2004) or Wan *et al.* (2009) or thin-layer asphalts by Synnefa *et al.* (2011). The latter reported surface temperature reductions ranging from 5°C to 15°C depending on the coating compared to standard asphalt concrete. Anting *et al.* (2017) tested the performance of NIR-reflective ceramic waste tiles which were up to 4° to 6°C cooler than a reference asphalt concrete tile.

By reducing their surface temperature, reflective pavements have been found to reduce air temperatures by up to 0.9°C for cities in California, results varying per city according to local conditions (Mohegh *et al.*, 2017). Taleghani *et al.* (2019) obtained comparable results in their study of reflective pavements applied to Los Angeles. Many more numerical studies of the phenomenon can be found in the literature, which report 1-2 m height air temperature cooling in the order of 1°C or less (Synnefa *et al.*, 2011; Battista and Pastore, 2017; Kolokotsa *et al.*, 2018; Kyriakodis and Santamouris, 2018).

One of the limits of reflective materials is surface aging and soiling, which cause performance degradation (Bretz and Akbari, 1997), though accelerated aging methods have been developed to determine this evolution ahead of time (Sleiman *et al.*, 2014; Takebayashi *et al.*, 2014).

This is an very important factor for pavement materials which are exposed to very intense urban pollution and to high levels of abrasive wear and soiling associated with road traffic (Rosado *et al.*, 2014; Kyriakodis and Santamouris, 2018; Lontorfos, Efthymiou and Santamouris, 2018; Tsoka *et al.*, 2018). In addition to encountering significant difficulties in obtaining high-albedo aggregates and in deploying high-albedo cement for cool asphalt and concrete pavement construction, Rosado *et al.* (2014) observed rapid albedo degradation due traffic in under 10 weeks. In their experiment, chip seals with high albedo aggregates had the most durable results compared to coatings. Furthermore, in cases when a surface layer is added onto a traditional pavement, mechanical issues may arise, limiting pavement lifespan (Sha et al., 2017).

Another concern in the application of reflective materials is their impact under winter conditions. Indeed, although it is only to a lesser extent, cooling observed in the summer also occurs in the winter, thus potentially increasing building heating demand. Recent work on cool roofs seems to indicate that the benefits in summer are not offset by winter penalties for cold climates (Akbari, 2014). More recently, thermochromic coatings, which change color depending on their temperature, are being developed and studied in the lab. These can be designed to increase their albedo above a certain temperature and remain dark below it. These coatings offer a chance to enjoy the cooling properties of high albedo in the summer without affecting energy demand in the winter (Santamouris, Synnefa and Karlessi, 2011). These materials, which already face short lifespans for rooftop or façade applications, a handful of research teams have begun investigating their potential when added to asphalt binders (Hu and Yu, 2013).

Another avenue for future work on reflective pavements involves assessing and reducing their environmental impacts. Indeed, it has been found that in certain cities, cool coatings, chip seals or bonded concrete overlaid on asphalt provide fewer life cycle gains than penalties (Gilbert *et al.*, 2017).

Another challenge which has been identified is the risk that under certain conditions the increase in reflected radiation caused by high albedo materials can have net negative impacts on pedestrian thermal comfort, defeating their purpose altogether (Erell *et al.*, 2014). In their findings, Erell *et al.* (2014) find that it is high-albedo building façades under certain urban and insolation configurations which are the principal source of the increased stress for pedestrians.

The same phenomenon of street canyon multi-reflections limits the street-scale impact of diffuse reflective pavements as part of the energy reflected away by the reflective pavements is absorbed by the adjacent buildings. In principle, this is not a concern for retro-reflective pavements, which preferentially reflect incident radiation back in the same direction, i.e. back towards the Sun (Rossi *et al.*, 2016).

**Green and evaporative pavements**

Green and evaporative pavements do not rely on higher albedo to reach lower surface temperatures. In fact, evaporative pavements are often less reflective than their standard counterparts, particularly in the case of pervious pavements. Instead, they release absorbed solar heat in the form of latent heat thanks to the evaporation of liquid water. This is accomplished by the evaporation of water trapped in the pavement material, with (green) or without the help of a herbaceous vegetation layer and the evapotranspiration of water they provide during photosynthesis. In both cases, water must be present in the pavement and available for evaporation or evapotranspiration. The more water is available in the material, the longer the cooling effect will last.

Evaporative pavements often rely on pervious materials, with or without water retention. The terms porous, pervious and permeable are often used interchangeably to describe a material that allows water to flow through it. However, porosity is a measure of the proportion of voids in a given material, while its permeability is a measure of its ability to let a fluid pass through it.

Porous pavements are usually obtained by increasing the proportion of large aggregates while reducing the proportion of fine aggregates in the pavement mix. The binder can be cement-, asphalt or resin-based. The void content is typically in the 15% to 25% range. Due to the presence of these voids, the mechanical strength of pervious pavements is usually lower than that of equivalent impervious pavements, typically 4-15 MPa. Increasing the proportion of binder can compensate for this with compressive strengths of up to 32 MPa, with limited reductions in permeability (Kevern, Schaefer and Wang, 2009). In addition, thicker layers of pervious pavement can be used to reach the same resistance.

Permeability can be measured in the field using the ASTM C1701 test method based on the constant head principle for pervious concrete pavements, while ASTM PS129-01 is used for pervious asphalt pavements [refASTMC1701].

To maintain their permeability, pervious pavements may require cleaning by vacuum sweeping, particularly in low speed traffic areas, while pervious highway pavements are considered to clean themselves with high-speed traffic. However, this may not be a problem with regards to lowering evaporative pavement surface temperature as clogging tends to promote capillarity to a certain degree. However, good drainage, principally ensured by sufficient pore size, is necessary to avoid issues with freeze-thaw in winter.

When dry, the air included in the voids of pervious pavements acts as a thermal insulator, increasing its surface temperature amplitude while dampening the temperature signal in the base layers. This is positive in winter, resulting in fewer occurrences of freezing temperatures under the surface course as observed by Kevern, Schaefer and Wang (2009) with cement concrete pavements. However, under hot weather conditions, this can lead to warmer surface temperatures for dry pervious pavements than for standard ones, i.e. a negative impact for urban climate (Kevern, Schaefer and Wang, 2009; Li, Harvey and Jones, 2013; Hassn et al., 2016).

Pervious pavements are often promoted as efficient tools to reduce stormwater runoff and improve runoff water quality, in addition to reducing rolling noise from traffic and improving road safety by providing a drier surface during rain events. However, while pervious materials are very useful in offsetting rainwater runoff into sewer systems, they are not necessarily very efficient at providing evaporative cooling if they do not include a mechanism for water retention. Indeed, only water stored close enough to the surface will evaporate rapidly due to thermal and aeraulic resistance of the porous material (Takebayashi and Moriyama, 2012; Nemirovsky, Welker and Lee, 2013).

Water-retaining materials make up for this by storing water close to the surface. This can be done thanks to capillary action which keeps moisture present at or near the pavement surface. This can be achieved by grouting the pores in pervious asphalt with a cement-based material that remains porous but with higher capillarity. This property has been clearly demonstrated by several studies which compare water-retaining pavements with porous ones, the former reaching surface temperatures up to 13°C lower than the latter (Kubo, Kido and Ito, 2006). Water-retaining pavers can also be created by sealing the bottom and sides of pervious concrete slabs, as studied experimentally on individual pavers by Qin et al. (2018). Storing approximately 9.5 mm of water, i.e. 9.5 L/m$^2$, the water-retaining block paver is found to maintain cooling for approximately 30h as opposed to its pervious and impervious counterparts, which provide only significant cooling immediately after sprinkling. This performance is reached despite the fact that the water-retaining paver has lower albedo, i.e. approximately 0.22 vs. 0.35-0.40, respectively. This evaporative cooling resulted in surface temperatures 2-10°C cooler than pervious and impervious pavers. When controlling for the difference in albedo by painting all surfaces black, this cooling effect is prolonged to 60h.

A number of studies of the potential for evaporative pavements to cool cities can be found in the literature (Asaeda, Ca and Wake, 1996; Kinouchi and Kanda, 1998; Asaeda and Ca, 2000; Kubo, Kido and Ito, 2006; Yamagata et al., 2008; Nakayama and Fujita, 2010; Kim et al., 2012; Takebayashi and Moriyama, 2012; Li, Harvey and Jones, 2013). These indicate that surface temperatures can be reduced in the order of 5-15°C compared to standard impervious asphalt concrete, depending on the thermal properties of the dry pavement. The best and longest performance is obtained with water-retaining pavements.

In addition, several authors have studied active evaporative cooling techniques of standard pavement structures, often referred to as pavement-watering or sprinkling (Kinouchi and Kanda, 1997; Takahashi et al., 2010; Bouvier, Brunner and Aimé, 2013; Hendel et al., 2014, 2015a, 2015b, 2016, 2018; Maillard et al., 2014; Hendel and Royon, 2015). Hendel et al. (2014, 2016), relying on street-cleaning trucks to sprinkle water in their experiment, report surface temperature reductions of 10-15°C during pavement insolation, resulting in air temperature reductions of up to 1°C. Fixed infrastructures, such as used in Japan by Kinouchi and Kanda (1997) and Takahashi et al. (2010), could be included into the design of future pavement retrofits in strategic urban areas.

Green materials are not generally included among cool materials, although they do provide lower surface temperatures compared to their standard counterparts. Few green pavements are available, though structures known as grid pavements can be found with grass-planted soil in the interstices or reinforced turf. Similar structures can be obtained with pavers by simply leaving openings between them. This can be as simple as increasing the joint width between pavers, as has been adopted in a number of cities such as Paris, France.

A thorough comparison of the cooling effects of 37 different kinds of green pavements has been conducted by Takebayashi and Moriyama (2009) in a parking lot in Kobe City, Japan. The authors report surface temperatures of up to 25°C cooler than standard asphalt concrete parking spaces. The solar reflectance of these structures ranges from 0.165 to 0.246, illustrating the significant cooling impact of evapotranspiration. Similarly, grass lawns generally exhibit surface temperatures approximately 20°C cooler than standard pavements (Asaeda and Ca, 2000; Takebayashi and Moriyama, 2012).

Grass lawns, which could be considered to represent an extreme of the spectrum of green pavement materials, generally offer the best performance with up to 20°C in surface temperature reduction compared to standard pavements. However, it should be noted that this performance is entirely cancelled out once the grass is dry and has died out.

Generally-speaking, green pavements are not suited to high mechanical stress applications such as high traffic and will not fare well in areas that require snow plowing or the use of de-icing salts in winter.

**High inertia pavements : phase-changing pavements**

To reduce their surface temperatures, high inertia pavements have neither higher reflectivity to limit solar heat absorption nor porosity to release absorbed heat by evaporating water. Instead, they attempt to offer a higher-than-normal thermal inertia. It is quite difficult to increase the inertia of concrete or asphalt concrete by changing formulations, e.g. aggregates. However, it is possible to add phase-change material (PCM) in their mix, which can be chosen or designed to change phase at a desired temperature. The inclusion of PCMs aims to increase the thermal mass $C$ of the pavement surface layer, thus dampening the rate of temperature change as per equation (10).

From a thermodynamics point-of-view, the phase change of a pure substance takes place at constant temperature and pressure. In other words, a pure substance solid melts into a liquid at constant temperature, while absorbing heat. If the process is reversed, resolidifying the obtained liquid releases exactly the same amount of heat at the same constant temperature. This is the same process when evaporating water from a liquid to a gas or melting water ice to a liquid. However, changing from a liquid to a gas is impractical as the specific volume of the substance changes very significantly. This is not the case when changing from a solid to a liquid which are both condensed forms of matter. PCMs therefore focus on the latter phase change and are also called solid-liquid PCMs.

Contrary to the evaporation of water in green and evaporative pavements, PCMs are trapped inside the pavement. Therefore, as the pavement cools down at night, the melted PCMs will resolidify, releasing the same amount of heat absorbed during the day. The amount of energy stored and released by the pavement is therefore not modified by the inclusion of PCMs. However, they do lower the rate of sensible heat exchange with the atmosphere.

Effectively, adding PCMs to a pavement increases its thermal inertia. While they absorb heat during the day, the temperature of phase-changing pavements will rise more slowly than traditional pavements. At night, the process is reversed: their temperature will reduce more slowly as they release the heat absorbed during the previous day. On average, this results in temperatures lower than standard pavements during the day, but higher at night.

As such, phase-changing pavements will likely not be effective means of limiting the UHI effect, since they will tend to exacerbate the reduced rate of nocturnal temperature decrease in cities. However, they are effective at limiting urban heat, i.e. limiting maximum temperatures during the day.

Typical PCMs are organic (bio- or petroleum-based paraffins, carbohydrates or lipids), inorganic salt hydrates or eutectics, i.e. not a pure substance. The principal design characteristics of PCMs are their melting temperature, latent heat of fusion, cyclability and thermal conductivity. Melting temperature determines the temperature at which phase change occurs and should fall in the range of temperatures reached within the pavement during the climate cycle being addressed. Latent heat of fusion is a measure of the amount of heat absorbed to fully melt the PCM or released to fully solidify it, desirable values are generally considered to be 200 kJ/kg or higher. Cyclability describes the number of fusion-solidification cycles that a PCM can undergo before its performance degrades. Finally, a PCM's solid-phase thermal conductivity is an important parameter in order to ensure rapid heat conduction into the solid PCM to facilitate melting.

PCMs can be included in materials directly, although this is generally not recommended as they may flow through the material's pores when in liquid phase. This may also pose issues with mechanical strength, specifically for concrete (Hunger *et al.*, 2009). PCMs are often microencapsulated to prevent this, but this is not resistant to hot mix asphalt temperatures. To solve this issue, PCMs can be included in light-weight aggregates used in asphalt concrete mixtures, as detailed by Ryms *et al.* (2015).

Several studies focus on the use of PCMs to reduce freeze-thaw in concrete and asphalt pavements (Sakulich and Bentz, 2012; Farnam *et al.*, 2017; Esmaeeli *et al.*, 2018; Yeon and Kim, 2018; Mahedi, Cetin and Cetin, 2019; Nayak, Krishnan and Das, 2019). In addition, some studies focus on the use of PCMs to reduce concrete curing temperatures or asphalt setting temperatures (Si *et al.*, 2015; Arora, Sant and Neithalath, 2017; She *et al.*, 2019).

However, relatively few articles focus on their use to mitigate urban heat (Chang, Li and Chang, 2007; Bo Guan, Biao Ma and Fang Qin, 2011; Chen *et al.*, 2011; Qin, 2015; Ryms *et al.*, 2015; Ryms, Denda and Jaskuła, 2017). In the lab, Chen *et al.* (2011) observed temperature reductions of 2°C for PCM pavement samples. Surface temperature reductions of about 5° were reported by Ryms *et al.* (2015) and Ryms, Denda and Jaskuła (2017) for PCM asphalt concrete containing 3% PCM by total weight. Athukorallage *et al.* (2018) investigated the use of adding a PCM layer beneath the surface course in asphalt concrete pavement structures. This was found to be counterproductive due to the low thermal conductivity of PCMs compared to pure asphalt concrete. However, a maximum surface temperature reduction of 4°C was found if the surface course includes an optimum 30% PCM by volume.

These surface temperature cooling effects are significantly lower than what can be obtained with reflective or green and evaporative pavements. Furthermore, adding PCM to concrete and asphalt concrete pavements poses challenges in terms of workability, mechanical strength and PCM inclusion.

Further work is necessary to achieve sufficient surface temperature cooling for phase changing pavements to merit application and become competitive with reflective or green and evaporative pavements.

**High conduction and heat-harvesting pavements**

High conduction and heat-harvesting pavements both increase the surface conduction term in equation (10). They reach this goal either by increasing the thermal conductivity $\lambda$ of the pavement structure, or by increasing the temperature gradient by reducing temperatures below the surface layer, typically via a heat exchanger. In the latter case, the heat exchanged can be harvested and put to productive use. Overall, high conduction pavements have been studied to a lesser extent than heat-harvesting pavements.

A number of different approaches have been proposed to increase the conductivity of pavements which principally determined by the conductivity of the binder and aggregates, as well as the void content. In certain cases, highly conductive powder can be added such as graphite. Indeed, Yinfei, Qin and Shengyue (2014) found that adding low conductivity material to the surface layer of asphalt concrete pavement increases surface temperature, as opposed to the addition of graphite which had the opposite effect. This design was studied with the aim of reducing temperatures deep inside the pavement, e.g. to limit permafrost thawing or rutting. Yinfei and Shengyue (2015) further observed surface cooling in the order of 1°C following the addition of vertical steel rods in the mid-course of an asphalt concrete pavement structure with the same low conductivity surface course. An improved version of the steel rod pavement was found to reduce surface temperatures by up to 3.5°C without increasing rutting (Yinfei *et al.*, 2018). Even better daytime surface temperature reductions can be expected if the conductivity of the surface course is increased instead of being decreased.

Heat-harvesting pavements belong to a larger category of energy-harvesting pavements. These include pavements which produce electricity such as photovoltaic pavements, thermoelectric and piezoelectric pavements, which respectively harvest sunlight, heat and traffic-induced mechanical vibrations, and heat-harvesting pavements, which directly harvest heat from the pavement structure. Heat-harvesting and photovoltaic (PV) pavements can also be referred to as solar pavements or solar collector pavements (Wang, Jasim and Chen, 2018; Papadimitriou, Psomopoulos and Kehagia, 2019).

The energy produced by solar pavements creates an additional outbound flow $\dot{q}$ not included in equation (10). By adding this term to the right-hand side of equation (10), we obtain:

$$S + L = \rho C \frac{\partial T_S}{\partial t} dz + \alpha S + (1 - \epsilon)L + \epsilon \sigma T_S^4 + h(T_{air} - T_S) + lE - \lambda \frac{\partial T_z}{\partial z}\bigg|_{z=dz} + \dot{q} \qquad (11)$$

As can be seen, the energy produced by solar collector pavements directly contributes to reducing its temperature, provided that solar pavement design does not significantly affect the other pre-existing outbound flows.

Noting $\eta$ as the overall efficiency of the solar pavement and its components, $\dot{q}$ is expressed by:

$$\dot{q} = \eta S \qquad (12)$$

PV pavements are included in this section even though they do not actually harvest heat from the pavement but sunlight, which is the source of the heat harvested by the other solar pavement designs.

**PV pavements**

PV pavements consist of a PV panel and its electronics, reinforced to bear traffic loads, placed on the surface layer of a standard pavement structure. Based on the photoelectric effect, the PV panels turn sunlight into electricity.

Few research papers are available on the energy efficiency and surface cooling potential of PV pavements (Efthymiou *et al.*, 2016; Dezfooli *et al.*, 2017; Papadimitriou, Psomopoulos and Kehagia, 2019). The energy efficiency of PV pavements, due to the addition of a reinforced glass layer on top of the solar cells, is typically lower than standard PV panels or cells. Efthymiou *et al.* (2016) find a 2% drop in energy efficiency due to the reinforced glass surface of the photovoltaic pavement, with an initial panel efficiency of 14%. Dezfooli *et al.* (2017) report that photovoltaic pavement efficiency is half that of the reference solar cell, however its initial efficiency is very low (1,1%). Xiang *et al.* (2018) indicate PV pavement efficiencies in the order of 9% due to the negative impact of high pavement temperatures. They propose a hybrid PV-thermal (PVT) approach to improve electrical efficiency to 11.5% and reach a combined thermal and electrical efficiency of between 45% and 67%.

In terms of surface temperature cooling, Efthymiou *et al.* (2016) report a reduction of up to 8°C for PV pavements compared to standard asphalt concrete.

**Thermoelectric pavements**

To turn heat into energy, thermoelectric pavements rely on the Seebeck effect, which can produce an electric field from a temperature gradient between the hot and cold junctions of a thermoelectric generator (TEG). TEGs can either be embedded in the pavement and exploit the pavement temperature gradient directly or they can combined with a pipe-pavement for use with pavement-heated and cold water (Guo and Lu, 2017; Wang, Jasim and Chen, 2018).

Few experimental studies of TEG pavements exist under realistic insolation conditions. However Jiang *et al.* (2017, 2018) studied a TEG system including horizontal aluminum heat conductors connected to TEGs placed on the side of the pavement structure, making use of a temperature difference of up to 10-20°C between the pavement and the ambient air. With 30x30cm asphalt test slabs placed outdoors, they obtained up to 840 J during a 7.5 hour insolation period in October, i.e. approximately 2.6 Wh/m$^2$.

Few experimental studies of TEG pavements exist under realistic insolation conditions. However Jiang *et al.* (2017, 2018) studied a TEG system including horizontal aluminum heat conductors connected to TEGs placed on the side of the pavement structure, making use of a temperature difference of up to 10-20°C between the pavement and the ambient air. The authors report the production of up to 33 Wh/m$^2$. In the best summertime conditions, with temperature differences in the order of 20-30°C, this is expected to reach 16 Wh/m$^2$ over 8 hours. This value is rather low, representing an average outbound heat flow of 2 W during the insolation period. This is explained by the TEG's low efficiency, i.e. approximately 5% of the power available from the temperature difference.

In terms of pavement cooling, the TEG pavement design exhibited surface temperatures up to 10°C cooler than traditional asphalt concrete in summer. This temperature reduction is misleading and should not be considered out of context. Although this value is indeed reached in the described experiment, only 5% of the extracted heat is turned into electricity, while the rest is released to the atmosphere.

**Heat-exchanger pavements**

The most studied solar pavements are those including a heat exchanging layer (van Bijsterveld *et al.*, 2007; Mallick, Chen and Bhowmick, 2009; Gao *et al.*, 2010; Bobes-Jesus *et al.*, 2013; Pascual-Muñoz *et al.*, 2013, 2014; García and Partl, 2014; Nasir, Hughes and Calautit, 2015, 2017b, 2017a; Pan *et al.*, 2015; Zhou *et al.*, 2015; Chiarelli, Dawson and García, 2015, 2017a, 2017b; Guldentops *et al.*, 2016; Alonso-Estébanez *et al.*, 2017; Chiarelli *et al.*, 2017). Embedded pavement pipes or a pervious pavement layer are used for this purpose, with either water or air circulating as the heat-transfer fluid. Harvested heat in summer can be used to supply hot water for nearby buildings or can be stored underground for use in winter for snow-melting of the heat-exchanger pavement.

Embedded-pipe heat-harvesting pavement performance in terms of surface temperature reduction has not been thoroughly studied to date. Zhou *et al.* (2015) report surface temperature reductions of 3°C to 6°C during the a summer day. Using air as the heat-transfer fluid, Chiarelli, Dawson and García (2017a, 2017b) report pavement cooling of up to 6°C.

It is clear however that design parameters such as fluid temperature, fluid flow rate, pipe depth and distance between pipes have a strong influence on surface temperature (van Bijsterveld *et al.*, 2007; Mallick, Chen and Bhowmick, 2012; Bobes-Jesus *et al.*, 2013; Pan *et al.*, 2015; Nasir, Hughes and Calautit, 2017a). Embedded pipes pose a certain number of engineering challenges in terms of stress and corrosion resistance as well as maintenance. Typical solar energy efficiency with embedded pipes is in the order of 25-40% (Gao *et al.*, 2010; Zhou *et al.*, 2015; Guldentops *et al.*, 2016).

Alternative approaches discard the need for pipes altogether by relying instead on a pervious sub-layer in which the heat-transfer fluid flows (Pascual-Muñoz *et al.*, 2013). Temperature reductions are more uniform with this approach and energy efficiencies of up to 85%.

Some systems are coupled with geothermal energy systems to store excess heat in summer for use in winter, including pavement deicing (Wang, Jasim and Chen, 2018).

**Combined cool pavement designs**

Recent work has outlined the improved impact of combined UHI mitigation techniques (Mohajerani, Bakaric and Jeffrey-Bailey, 2017). One such example has been studied by Higashiyama *et al.* (2016) who combine porous asphalt concrete with a light colored and absorbent sealer based on a ceramic powder made from waste. Although its impact on albedo was not reported, the authors report surface temperature reductions from 9°C to 20°C compared to standard porous asphalt concrete for the dry pavements, i.e. without rain or sprinkling.

Other venues can also be explored though they usually require a tradeoff, e.g. reflective or evaporative solar pavements which may offer even cooler surface temperatures in exchange for reduced harvesting efficiency. Another example, from Guo and Hendel (2018), concerns heat, or rather coolth, harvesting from a pavement-watering infrastructure.

The multitude of potential combinations offers many future research possibilities.